\documentclass[letter]{aa} 

\usepackage{graphicx}
\usepackage[varg]{txfonts}
\usepackage{natbib}
\usepackage{hyperref}

\begin{document}

 \title{A non-equilibrium ortho-to-para ratio of water in the Orion PDR\thanks{{\it Herschel} is an ESA space observatory with science instruments
       provided by European-led Principal Investigator consortia and with important participation from NASA.}}

 \author{Y. Choi\inst{1,2},
         F. F. S. van der Tak\inst{2,1},
         E. A. Bergin\inst{3},
         R. Plume\inst{4}}

   \institute{Kapteyn Astronomical Institute, University of Groningen, P.O. Box 800, 9700 AV, Groningen, The Netherlands \\
              \email{y.choi@astro.rug.nl} \and
              SRON Netherlands Institute for Space Research, P.O. Box 800, 9700 AV, Groningen, The Netherlands \and
              Department of Astronomy, University of Michigan, 500 Church Street, Ann Arbor, MI 48109, USA \and
              Department of Physics and Astronomy, University of Calgary, 2500 University Drive NW, Calgary, AB T2N 1N4, Canada}



  \abstract
   {The ortho-to-para ratio (OPR) of H$_2$O is thought to be sensitive to the temperature of water formation.
The OPR of H$_2$O is thus useful to study the formation mechanism of water.}
   {We investigate the OPR of water in the Orion PDR (Photon-dominated region), at the Orion Bar and Orion S positions,
using data from {\it Herschel}/HIFI.}
   {We detect the ground-state lines of ortho- and para-H$_2$$^{18}$O
in the Orion Bar and Orion S and we estimate the column densities using LTE and non-LTE methods.}
   {Based on our calculations, the ortho-to-para ratio (OPR) in the Orion Bar is 0.1 $-$ 0.5, which is unexpectedly low given the gas temperature of $\sim$ 85 K,
   and also lower than the values measured for other interstellar clouds and protoplanetary disks.
   Toward Orion S, our OPR estimate is below 2.}
   {This low OPR at 2 positions in the Orion PDR is inconsistent with gas phase formation and with thermal evaporation from dust grains,
but it may be explained by photodesorption.}

   \authorrunning{Y. Choi et al.}
   \titlerunning{The ortho-to-para ratio of water in the Orion PDR}

   \maketitle
%

\section{Introduction}
\label{intro}
Water is an important reservoir of interstellar oxygen and therefore a key ingredient in the chemistry of oxygen-bearing molecules.
In the interstellar medium, water can be formed by three different mechanisms.
In cold molecular clouds, water may be formed in the gas phase by ion-molecule chemistry,
through dissociative recombination of H$_3$O$^+$. In cold and dense cores, on the surfaces of cold dust grains,
O and H atoms may combine to form water-rich ice mantles.
These mantles will evaporate when the grains are heated to $\sim$ 100 K by protostellar radiation or sputtered by outflow shocks.
Third, in gas with temperatures above 300 K, reactions of O and OH with H$_2$ drive all gas-phase oxygen into water.
Such high temperatures may occur very close to the star due to heating by protostellar radiation, or outflow shocks
\citep[see][for a review]{vDishoeck13}.

The hydrogen atom carries a nuclear spin angular momentum and two species of molecular hydrogen exist (ortho-H$_2$ and para-H$_2$).
Other molecules with two or more hydrogen atoms also exhibit this characteristic with two independent spin isomers
 (e.g., H$_2$O, NH$_3$, CH$_4$).
Because of a difference in the energy of the rotational ground state (34 K for ortho-H$_2$O, 0 K for para-H$_2$O)
the ratio of ortho to para water vapor is dependent on the gas temperature in thermal equilibrium.
Above 40 K the OPR would be the ratio of the spin statistical
weights of 3 and below this temperature there is an expectation that the formation of para-H$_2$O would be successively favored over ortho-H$_2$O leading to a reduced OPR \citep{Mumma87}.
Since inelastic collisions do not change the OPR, this ratio may provide clues to the formation mechanism of water.
The measured OPRs of H$_2$O is 2$-$3 in solar system comets \citep{Mumma11} and in Galactic interstellar clouds \citep{Lis13, Flagey13},
while an OPR of $\sim$1 was measured for water vapor in the TW Hya disk \citep{Hogerheijde11}.

\begin{table*}
\caption{Observed lines.}
\label{tab1}
\centering
\begin{tabular}{l l l c c c c c c c c}
\hline\hline
Source & Molecule & Transition & $\nu$ & $E_{\rm up}$ & $T_{\rm sys}$ & $t_{\rm int}$ & Beam      & $\eta_{\rm mb}$ & $\delta v $ & rms  \\
       &          &            & (GHz) & (K)          & (K)           & (min)         & (\arcsec) &                 & (MHz)       & (mK)   \\
\hline
Orion Bar & o-H$_2$$^{18}$O & $1_{10}-1_{01}$ & 547.676  & 60.5  & 63   & 88  & 38.7 & 0.75 & 0.48 & 2.67  \\
          & p-H$_2$$^{18}$O & $1_{11}-0_{00}$ & 1101.698 & 53.4  & 350  & 56  & 19.2 & 0.74 & 1.1  & 102   \\
\hline
Orion S & o-H$_2$$^{18}$O & $1_{10}-1_{01}$ &  547.676  & 60.5 & 71   & 7.4 & 38.7 & 0.75 & 1.1 & 26   \\
        & p-H$_2$$^{18}$O & $1_{11}-0_{00}$ &  1101.698 & 53.4 & 366  & 10  & 19.2 & 0.74 & 1.1 & 121  \\
\hline
\end{tabular}
\end{table*}

In this paper we will present measurements of the rotational emission of both spin isomers of water vapor towards the Orion photon-dominated region.
Photon-dominated regions (PDRs) are the surface regions of
molecular clouds, where ultraviolet radiation with photon energies between 6 and 13.6 eV drives the thermal and chemical balance of the gas \citep{Hollenbach99}.
Shielding of the UV radiation by dust and gas creates a layered structure where a sequence of
different chemical transitions is produced by the gradual attenuation of the UV field \citep{Ossenkopf07}.

The Orion Molecular Cloud 1 (OMC-1), at a distance of $\sim$ 420 pc \citep{Menten07},
is one of the nearest massive star forming regions.
Parts of the OMC-1 region are ionized by the Trapezium cluster creating an HII region.
The Orion Bar PDR stands out as a ridge to the Southeast of the Trapezium cluster.
Observations at infrared and submillimeter wavelengths
indicate a geometry for the Bar where the PDR is wrapped around the HII region created by the Trapezium stars, and changes from a face-on
to an edge-on view where the molecular emission peaks \citep{Hogerheijde95}.
The mean density of the Bar is about 10$^{5}$ cm$^{-3}$, and the gas temperature is 85 K in the interior, rising to $\sim$ 150 K at the PDR surface \citep{Larsson03}.
The impinging radiation field is (1$-$4) $\times$ 10$^{4}$ $\chi_{0}$,
where the Draine field $\chi_{0}$ = 2.7 $\times$ 10$^{-3}$ erg s$^{-1}$ cm$^{-2}$ \citep{Draine78}.
The clumpiness of the PDR inferred by \citet{Hogerheijde95} is confirmed by interferometric data
\citep{Lis03}, with densities of up to 1.5 $\times$ 10$^6$ to 6 $\times$ 10$^6$ cm$^{-3}$.
In contrast the densities of the interclump medium fall between a few 10$^4$ cm$^{-3}$ \citep{YoungOwl00}
and 2 $\times$ 10$^5$ cm$^{-3}$ \citep{Simon97}.

Orion S is an active
star-forming region, located 1$\arcmin$ southwest of the Trapezium, as indicated by the number of outflows and Herbig-Haro flows \citep{Zapata06}.
The mass of Orion S is $\sim$100 M$_\sun$ and
the size of this region is similar
to that of Orion BN/KL, but its bolometric luminosity of 10$^4$ $L_{\rm \sun}$ is
an order of magnitude lower \citep{Mezger90}, which may indicate that Orion S is less evolved \citep{McMullin93}.
The UV radiation field is estimated to be $\chi$ $\sim$1.5 $\times$ 10$^5$$\chi_{0}$ at the position of Orion S \citep{Herrmann97}, about a factor of 10 higher than that in the Orion Bar.
Due to the irradiation by the nearby Trapezium cluster, the
part of the Orion S region facing the Trapezium cluster includes an ionization front and a face-on PDR.

This paper uses {\it Herschel}/HIFI observations of water lines in the Orion PDR, at the Orion Bar and Orion S positions.
With its much higher spatial and spectral resolution and higher sensitivity than previous space missions,
we investigate the ortho-to-para ratio of water, providing new information on the formation mechanism of water in these regions.


\begin{table*}
\caption{Line parameters obtained from Gaussian fits.}
\label{tab2}
\centering
\begin{tabular}{l l l c c c c c}
\hline\hline
Source & Molecule & Transition & $\int T_{\rm MB}dV$ & $V_{\rm LSR}$ & $\Delta V$    & $T_{\rm MB}$ \\
       &          &            & (K km s$^{-1}$)     & (km s$^{-1}$) & (km s$^{-1}$) & (K)           \\
\hline
Orion Bar & o-H$_2$$^{18}$O & $1_{10}-1_{01}$ & 0.23 (0.01) & 10.22 (0.01) & 1.93 (0.03) & 0.11  \\
          & p-H$_2$$^{18}$O & $1_{11}-0_{00}$ & 0.56 (0.12) & 9.91  (0.19) & 1.84 (0.48) & 0.29  \\
\hline
Orion S & o-H$_2$$^{18}$O & $1_{10}-1_{01}$ & 1.17 (0.06) & 7.45 (0.13) & 4.66 (0.27) & 0.23  \\
        & p-H$_2$$^{18}$O & $1_{11}-0_{00}$ & 3.48 (0.27) & 7.05 (0.18) & 5.53 (0.64) & -0.59\tablefootmark{a}  \\
\hline
\end{tabular}
\tablefoot{\tablefoottext{a}{Absorption line detected in Orion S}}
\end{table*}


\section{Observations}
\label{obs}
The CO$^{+}$ peak in the Orion Bar was observed with the Heterodyne Instrument for the Far-Infrared
\citep[HIFI,][]{deGraauw10} onboard ESA's {\it Herschel} Space Observatory \citep{Pilbratt10},
   in all HIFI bands as part of
   the {\it Herschel} observations of EXtra-Ordinary Sources (HEXOS) guaranteed-time key program \citep{Bergin10}.
   The coordinates of the observed position of the CO$^{+}$ peak in the Orion Bar are 05$^{\rm h}$35$^{\rm m}$20$^{\rm s}$.6
   and -05$\degr$25$\arcmin$14$\arcsec$ (J2000).

In this paper we use the p-H$_2$$^{18}$O $1_{11}-0_{00}$ line from the HIFI band 4b spectral line survey.
This observation was carried out in April 2011 in load chop
mode with a redundancy of 4 and with a total integration time of 0.7 h.
The Wide-Band Spectrometer (WBS) backend was used which covers 4 GHz
bandwidth in four 1140 MHz subbands at 1.1 MHz resolution.
In addition to the HIFI spectral scan, the o-H$_2$$^{18}$O $1_{10}-1_{01}$ line was observed in September 2010
as a deep integration with a total integration time of 1.5 h in frequency switch mode.

Orion S was observed with a complete HIFI spectral scan as part of the HEXOS program.
The observations were pointed toward 05$^{\rm h}$35$^{\rm m}$13$^{\rm s}$.4
and -05$\degr$24$\arcmin$08$\arcsec$.1 (J2000).
We use data from HIFI bands 1a (o-H$_2$$^{18}$O $1_{10}-1_{01}$) and 4b (p-H$_2$$^{18}$O $1_{11}-0_{00}$).
The scans were observed using dual beam switch (DBS) observing mode. The WBS backend with a 1.1 MHz resolution was used.

 We performed calibration of the data, removal of standing waves and
 spurs and sideband deconvolution using the {\it Herschel} Interactive
Processing Environment \citep[HIPE,][]{Ott10} version 10.0.
Further analysis was done by the CLASS\footnote{http://www.iram.fr/IRAMFR/GILDAS/} package.

The frequencies, energy of the upper levels, system temperatures, integration times, and rms noise level
at a given spectral resolution for each of the lines are provided in Table~\ref{tab1}
along with the beam sizes and main beam efficiencies from \citet{Roelfsema12}.


\section{Results}
\label{results}

The HIFI spectra of the Orion Bar show pure single-peaked emission profiles in the ground-state lines of para- and ortho-H$_2$$^{18}$O (Fig.~\ref{fig1}).
In contrast, in Orion S, the ground-state line of ortho-H$_2$$^{18}$O appears in emission
but the ground-state line of para-H$_2$$^{18}$O is detected in absorption (Fig.~\ref{fig2}).
This effect may be due to the stronger continuum in Orion S
than in the Orion Bar, which increases towards higher frequencies assuming the dust and gas are well mixed (continuum level $\sim$ 4.6 K at 1101.7 GHz in Orion S compared to 0.3 K for the Bar). In addition, LVG models by \citet{Cernicharo06} predict that ortho- and para-H$_2$O lines appear in absorption or emission depending on the adopted conditions.

We extract line parameters from the observed profiles by fitting Gaussians.
The o-H$_2$$^{18}$O $1_{10}-1_{01}$ in Orion S
has a hint of a self-reversal, but this is unfortunately at the level of noise in the data and the parameters of this line were determined by fitting one Gaussian;
Table~\ref{tab2} gives the results for all lines.
The components in the Orion Bar show
similar line profiles, $\Delta${\it V} $\sim$ 1.8 $-$ 1.9 km s$^{-1}$ and $V_{\rm LSR}$ $\sim$10 km s$^{-1}$,
suggesting that these two lines originate in the same gas.
These parameters are also similar to those of CO, H$_2$CO and other dense gas tracers \citep{Leurini06}.
On the other hand, the line profiles in Orion S have a width of 4.6 $-$ 5.5, which is broader than in the Orion Bar, and a velocity of $\sim$ 7.0 $-$ 7.4 km s$^{-1}$.
This velocity shift is commensurate with the known N-S velocity gradient seen along the Orion Molecular Ridge
\citep{Ungerechts97}.
The observed line widths and LSR velocities are similar to those of CO isotopologues \citep{Peng12}.

\section{Analysis}
\label{analysis}
\subsection{Orion Bar}
\label{orionbar}
We estimate column densities assuming LTE and also explored models where the populations are not in LTE.
For the LTE calculations, we have some additional supporting evidence based on other observations of both sources obtained as part of the HEXOS program.
(1) The ground state lines of H$_2$$^{17}$O are not detected (rms $\sim$ 0.04 K at the o-H$_2$$^{17}$O $1_{10}-1_{01}$ line
and rms $\sim$ 0.2 K at the p-H$_2$$^{17}$O $1_{11}-0_{00}$ line) which limits the optical depth of the H$_2$$^{18}$O lines to be below 0.7.
In the following we assume that the emission is optically thin.
(2) The water emission is not arising from very warm ($>$ 100 K) and dense gas ($>$ 10$^8$ cm$^{-3}$)
towards either the Orion Bar or Orion S as we do not detect emission arising
from excited states (p-H$_2$$^{18}$O $2_{02}-1_{11}$, o-H$_2$$^{18}$O $2_{12}-1_{01}$, rms $\sim$ 0.2 $-$ 0.7 K),
which implies a limit on the excitation temperature of less than 100 K.
(3) The difference in beam size between the ortho-H$_2$$^{18}$O 547 GHz line and para-H$_2$$^{18}$O 1101 GHz line observations could lead us to underestimate the OPR by factors up to 4.
However this is unlikely as water emission in the Orion ridge extends over several arc-minutes \citep{Melnick11}.
Furthermore, the detected lines are likely to come from the same gas considering the velocities and line widths (see Table~\ref{tab2}).

We therefore derive the column densities of ortho- and para-H$_2$$^{18}$O in the Orion Bar for different excitation temperatures ($T_{\rm ex}$ = 50 $-$ 100 K)
and find values of $\sim$ 3.0 $\times$ 10$^{10}$ cm$^{-2}$ for the ortho-H$_2$$^{18}$O line and $\sim$ 1.0 $\times$ 10$^{11}$ cm$^{-2}$ for the para-H$_2$$^{18}$O line.
The derived column densities are not strongly sensitive to the assumed excitation temperature.
With the above assumptions, and in LTE, we derive an OPR of $\sim$ 0.3.

The ground state emission lines of ortho and para-water are not identical in their excitation characteristics.
Moreover both lines have high critical densities and we therefore explore non-LTE models of H$_2$O
using the RADEX code \citep{vdTak07} and state-of-the-art quantum mechanical collision rates of para- and ortho-H$_2$O with para- and ortho-H$_2$ \citep{Daniel11} as provided at the LAMDA database \citep{Schoier05}, assuming thermal values for the o/p ratio of H$_2$.
For this exploration we generate a grid of models with values of $T_{\rm kin}$ = 20, 60, and 100 K, values of $n$(H$_2$) = 10$^4$, 10$^6$, and 10$^8$ cm$^{-3}$, and fix the background radiation temperature at 2.73 K for the Orion Bar.
Within the range of assumed densities and temperatures we find that the analysis remains consistent with
a low OPR with values ranging from 0.1 to 0.5 (see Table~\ref{tab3}).
For the case where H$_2$O emission is arising from the warm surface of the PDR, \citep[e.g.][]{Hollenbach09},
we perform an additional solution with $T_{\rm gas}$ = 200 K.
For the density we use the detailed model of \citet{Nagy13} which has a density of 10$^5$ cm$^{-3}$ at 200 K; with this assumption the derived ortho-to-para ratio is 0.14.

As an alternative background radiation field, we adopt a modified blackbody distribution with
a dust temperature of $T_{\rm d}$ = 49 K and a dust emissivity index of $\beta$ = 1.6 for the interior of the Orion Bar \citep{Arab12},
so that the absolute dust opacity of $\tau_{\rm d}$ = 0.21 at 971 GHz.
This model predicts that the column densities are similar to those at the background radiation temperature of 2.73 K
under the same conditions of $T_{\rm kin}$ and $n$(H$_2$) and the ortho-to-para ratio of water is $\sim$ 0.1 $-$ 0.5 (Table~\ref{tab3}).

\subsection{Orion S}
\label{orions}
To estimate the column density for the absorption component of p-H$_2$$^{18}$O $1_{11}-0_{00}$, which is detected in Orion S,
we derived the optical depth using the expression
\begin{equation}
\tau=-{\rm ln}\left(\frac{T_{\rm line}}{T_{\rm cont}}\right)
\label{eq1}
\end{equation}
where $T_{\rm cont}$ is the single side band (SSB) continuum intensity
assuming that the continuum is completely covered by the absorbing layer
and $T_{\rm line}$ is the intensity at the absorption dip with continuum.
We apply a linear baseline fit in the vicinity of the absorption line to derive the continuum intensity ($\sim$ 4.6 K with uniform beam filling) at the absorption peak.
Deriving the optical depth from the line-to-continuum ratio is based on the assumption that
the excitation temperature is negligible with respect to the continuum temperature (i.e., no emission filling in the absorption)
and that the line is not saturated.

If all water molecules are in the para ground-state,
the velocity integrated absorption is related to the molecular column density by
\begin{equation}
N = \frac{8\pi\nu^{3}g_l}{c^{3}Ag_u}\int{\tau dV} \,,
\label{eq2}
\end{equation}
   where $N$ is the column density, $\nu$ the frequency, $c$ the speed of light, and $\tau$ is the optical depth.
   $A$ stands for the Einstein-A coefficient and $g_l$ and $g_u$ are the degeneracy of the lower and the upper
   level of the transition.
Integrating between $V$ = 0 and 13 km s$^{-1}$,
we find a column density of $\sim$ 2.0 $\times$ 10$^{12}$ cm$^{-2}$ for the para-H$_2$$^{18}$O line in Orion S.

Assuming LTE and using the same assumptions
as for the Bar given the limit on the optical depth ($<$ 0.7) by the non-detection of H$_2$$^{17}$O and the $T_{\rm ex}$ limit ($<$ 100 K)
from the excited states, we find a beam-averaged column density of $\sim$ 2.0 $\times$ 10$^{11}$ cm$^{-2}$ for the ortho-H$_2$$^{18}$O line.
Thus, assuming LTE, the ortho-to-para ratio of water is $\sim$ 0.1 in Orion S.

As before we perform a series of non-LTE calculations to confirm the low OPR derived for Orion S assuming LTE.
We therefore generate a grid of models with values of $T_{\rm kin}$ and $n$(H$_2$), and a background radiation field of 2.73 K.
We find an OPR $\sim$ 0.3 assuming $T_{\rm kin}$ = 100 K and $n$(H$_2$) = 10$^8$ cm$^{-3}$, but derive a value of 3
(or unphysically higher than 3) if $T_{\rm kin}$ = 60 K and $n$(H$_2$) = 10$^6$ cm$^{-3}$ or
for $T_{\rm kin}$ = 20 K and $n$(H$_2$) = 10$^4$ cm$^{-3}$.
Thus the derived ortho-to-para ratio of water is strongly sensitive to the assumed physical conditions (Table~\ref{tab3}).

Further constraints on the OPR in Orion S come from the non-detection of the ortho-H$_2$$^{18}$O $2_{12}-1_{01}$ line (1655.9 GHz)in our survey.
This line connects to the ortho-H$_2$O ground state and arises at a much higher frequency where the continuum level is higher
($\sim$ 6.1 K) than at the frequency of the para-H$_2$O ground state line.
This non-detection of the ortho-H$_2$$^{18}$O $2_{12}-1_{01}$ line in either emission or absorption
gives information on the limit of the ortho-H$_2$O column density in the absorbing gas.
Assuming that the ground-state line of ortho-H$_2$$^{18}$O $2_{12}-1_{01}$ appears in absorption,
we estimate its optical depth.
We adopt a line width of 5.5 km s$^{-1}$ from the para-H$_2$$^{18}$O $1_{11}-0_{00}$ line observations,
a continuum intensity of 6.1 K, and a rms of 0.4 K from the ortho-H$_2$$^{18}$O $2_{12}-1_{01}$ line observations.
Assuming that the ortho-to-para ratios of water are 3, 2.5, 2, and 1,
we derive a $T_{\rm line}$ of 4.8, 5.0, 5.2, and 5.6 K, respectively, using Eq.~(\ref{eq1}) and (\ref{eq2}).
If the ortho-to-para ratios of water are 3 and 2.5,
we should see absorption lines of ortho-H$_2$$^{18}$O $2_{12}-1_{01}$ with optical depth of 0.24 and 0.20, respectively.
However, for ortho-to-para ratios of 1 and 2, the ortho-H$_2$$^{18}$O $2_{12}-1_{01}$
would be seen in absorption with optical depth of 0.16 and 0.08 (within the noise), as illustrated in Fig.~\ref{fig3}.
We conclude that our data are consistent with OPR $\le$ 2 for cold water, but not with OPR $\ge$ 2.5.

\section{Discussion}
\label{disc}

Our derived OPR of H$_2$O in the Orion Bar is 0.1 $-$ 0.5 ($T_{\rm spin}$ $\sim$ 8 $-$ 12 K), which is well below that of other ISM sources and even lower than that toward TW Hya.
It is also much lower than expected based on the gas temperature of $\sim$ 85 K \citep{Hogerheijde95}.
Furthermore, the ortho-to-para ratio of water in Orion S is below 2 ($T_{\rm spin}$ $<$ 23 K).

Currently it is uncertain how low ortho-to-para ratios originate.
Gas phase formation via H$_3$O$^+$ dissociative recombination is expected to lead to an OPR of 3 due the fact that the reaction is exothermic
and the energy releases is well in excess of the ortho/para energy difference \citep{Hogerheijde11}.
It cannot originate from grain surfaces due to thermal evaporation as the measured dust temperature towards the Bar
(and presumably Orion S) is $\sim$ 35 $-$ 70 K \citep{Arab12} which is below the evaporation temperature of water ice \citep[$\sim$ 100 K;][]{Fraser01}

The low water OPR may be explained by photodesorption,
which has been argued to be the main formation mechanism for cold water vapor in the dense ISM \citep{Hollenbach09}.
In addition, a direct observational proof for photodesorption of H$_2$O is provided by the detection of gas-phase H$_2$O
towards the pre-stellar core L1544 \citep{Caselli12} and NGC 1333-IRAS4A protostar \citep{Mottram13}.
There are two possible processes for photodesorption, following photodissociation of H$_2$O ice into H and OH after absorption of a UV photon
\citep{Andersson08, Arasa10, Tielens13, vDishoeck13}:
1) H and OH recombine in the ice to form H$_2$O, which then has sufficient energy to desorb,
2) the energetic H atom kicks out a neighboring H$_2$O molecule from the ice, which is initiated by the same UV photon.
In option 1, the OPR should go to the statistical value of 3 because of the exothermicity of the reaction.
In option 2, the original OPR in the ice should be preserved. So if the grain temperature is low and the OPR equilibrated
to the grain temperature, the OPR should be low, in agreement with our results.
The relative importance of option 1 and 2 depends on the thickness of the ice layer and to a lesser extent on the ice temperature.
Roughly, they contribute about equally, but a detailed calculation is beyond the scope of this work.
The experiments by \citet{Yabushita09} show that the measured translational and rotational energies of H$_2$O ($v$ = 0) molecules photodesorbed
from amorphous solid water are in good agreement with those predicted by classical molecular dynamics calculations for the ``kick-out'' mechanism (option 2).

The ortho-to-para ratios of water have been measured in many different environments
but the {\it Herschel}/HIFI H$_2$O observations in the Orion PDR show an unusually low ortho-to-para ratio of water.
This opportunity will be further explored in a future paper,
where we will estimate ortho-to-para ratio of other molecules in the Orion PDR, and we will compare ortho-to-para ratio of water in other PDRs.
In addition, the {\it Herschel}/HIFI H$_2$O observations toward the Orion PDR show the structure in more detail than any previous study.
It will also be further study in a future paper, where we will estimate H$_2$O abundance profiles.





\begin{acknowledgements}
We thank the referee for the constructive suggestions. We also thank the editor Malcolm Walmsley for additional helpful comments.
The authors thank Ewine van Dishoeck, Edith Fayolle, Jean-Hugh Fillion, and Annemieke Petrignani for useful discussions on photodesorption.
HIFI has been designed and built by a consortium of institutes and university departments from across
Europe, Canada and the US under the leadership of SRON Netherlands Institute for Space Research, Groningen,
The Netherlands with major contributions from Germany, France and the US. Consortium members are:
Canada: CSA, U.Waterloo;
France: CESR, LAB, LERMA, IRAM;
Germany: KOSMA, MPIfR, MPS;
Ireland, NUI Maynooth;
Italy: ASI, IFSI-INAF, Arcetri-INAF;
Netherlands: SRON, TUD;
Poland: CAMK, CBK;
Spain: Observatorio Astron{\'o}mico Nacional (IGN), Centro de Astrobiolog{\'{\i}}a (CSIC-INTA);
Sweden: Chalmers University of Technology – MC2, RSS \& GARD, Onsala Space Observatory, Swedish National Space Board, Stockholm University – Stockholm Observatory;
Switzerland: ETH Z{\"u}rich, FHNW;
USA: Caltech, JPL, NHSC.
\end{acknowledgements}

\bibliographystyle{aa}    
\bibliography{OrionOPR_L_biblio}    

\Online
\begin{appendix}
\section{Spectra of the ground-state H$_2$$^{18}$O lines}
\begin{figure}
\centering
\includegraphics[width=8cm]{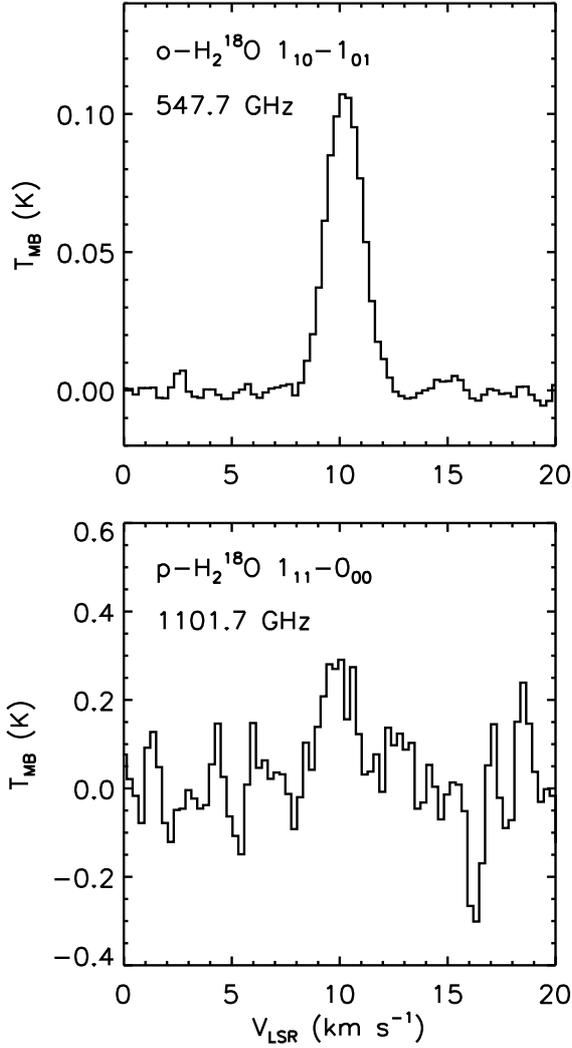}
      \caption{Spectra of the ground-state H$_2$$^{18}$O lines in the Orion Bar. The weak continuum has been subtracted.}
       \label{fig1}
\end{figure}

\begin{figure}
\centering
\includegraphics[width=8cm]{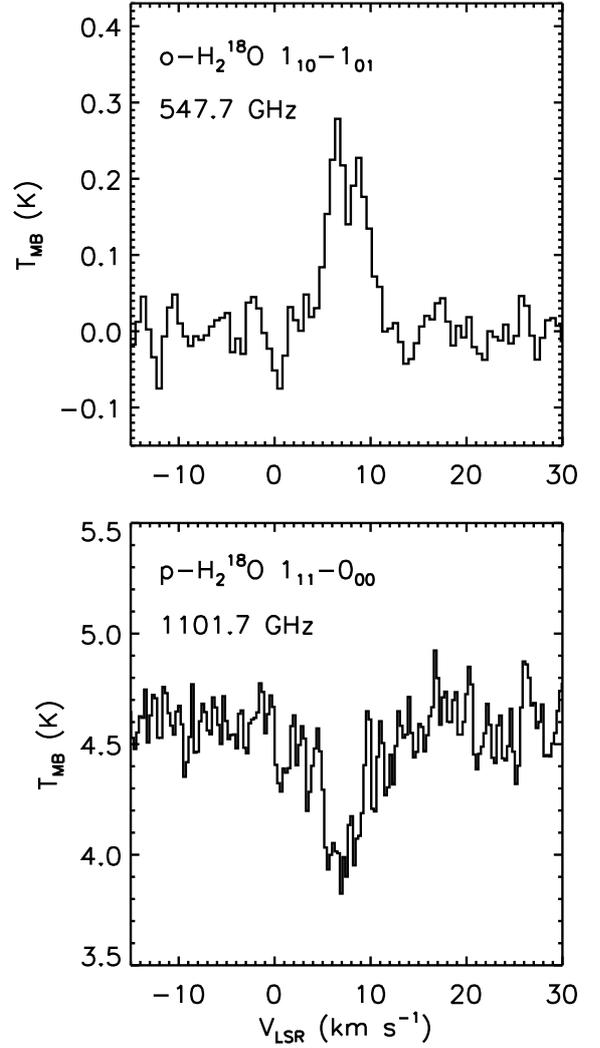}
\caption{Spectra of the ground-state H$_2$$^{18}$O lines toward Orion S. For the o-H$_2$$^{18}$O line, the continuum has been subtracted.
         The p-H$_2$$^{18}$O line shows absorption against the continuum.}
       \label{fig2}
\end{figure}

In Sect.~\ref{results},
we present the ground-state lines of ortho- and para-H$_2$$^{18}$O observed with {\it Herschel}/HIFI toward the Orion Bar (Fig.~\ref{fig1}) and Orion S (Fig.~\ref{fig2}).

\section{Column densities derived by non-LTE RADEX code}
In Sect.~\ref{analysis},
for the non-LTE calculations of ortho- and para-H$_2$$^{18}$O lines we generate a grid of models with values of $T_{\rm kin}$ = 20, 60, and 100 K, values of $n$(H$_2$) = 10$^4$, 10$^6$, and 10$^8$ cm$^{-3}$,
and fix the background radiation temperature at 2.73 K for the Orion Bar and Orion S using the RADEX code \citep{vdTak07}.
Table~\ref{tab3} presents the derived column densities for the adopted conditions from a full grid non-LTE calculations as examples.
The p-H$_2$$^{18}$O $1_{11}-0_{00}$ line in Orion S appears in absorption so we derive the column density using the optical depth (see Sect.~\ref{orions} for details).

As an additional model for the Orion Bar, for the background radiation field we adopt a modified blackbody distribution with
a dust temperature of $T_{\rm d}$ = 49 K and a dust emissivity index of $\beta$ = 1.6 by \citet{Arab12} for the interior of the Orion Bar, so that the absolute dust opacity of $\tau_{\rm d}$ = 0.21 at 971 GHz.
This RADEX model shows that ortho- and para-H$_2$$^{18}$O lines appears in absorption at the low density and low temperature (at $T_{\rm kin}$ = 20 K \& $n$(H$_2$) = 10$^{4}$ cm$^{-3}$), which is not consistent with our observations.



\begin{table*}
\caption{Examples of column densities for the adopted conditions from a full grid non-LTE calculations in the Orion Bar and Orion S.}
\label{tab3}
\centering
\begin{tabular}{l l l c c c c c}
\hline\hline
Source & Molecule & Transition & \multicolumn{3}{c}{$T_{\rm kin}$ \& $n$(H$_2$)  (K \& cm$^{-3}$)} \\
       &          &            & 20 \& 10$^{4}$ & 60 \& 10$^{6}$ & 100 \& 10$^{8}$                  \\
\hline
\multicolumn{6}{c}{$N$ from RADEX Model with $T_{\rm bg}$ = 2.73 K (cm$^{-2}$)} \\
\hline
Orion Bar & o-H$_2$$^{18}$O & $1_{10}-1_{01}$ & 2.19$\times$10$^{15}$ & 1.65$\times$10$^{12}$ & 1.20$\times$10$^{11}$ \\                                           
          & p-H$_2$$^{18}$O & $1_{11}-0_{00}$ & 3.08$\times$10$^{16}$ & 1.75$\times$10$^{13}$ & 2.66$\times$10$^{11}$  \\
\hline
Orion S & o-H$_2$$^{18}$O & $1_{10}-1_{01}$ & 1.13$\times$10$^{16}$ & 8.57$\times$10$^{12}$ & 6.06$\times$10$^{11}$ \\
\hline
\multicolumn{6}{c}{$N$ from RADEX Model with $T_{\rm d}$ = 49 K \& $\beta$ = 1.6 (cm$^{-2}$)} \\
\hline
Orion Bar & o-H$_2$$^{18}$O & $1_{10}-1_{01}$ & $-$ & 2.80$\times$10$^{12}$ & 1.49$\times$10$^{11}$ \\
          & p-H$_2$$^{18}$O & $1_{11}-0_{00}$ & $-$ & 2.01$\times$10$^{13}$ & 2.96$\times$10$^{11}$  \\
\hline
\end{tabular}
\end{table*}

\section{Further constraints on the OPR in Orion S}
\begin{figure}
\centering
\includegraphics[width=8cm]{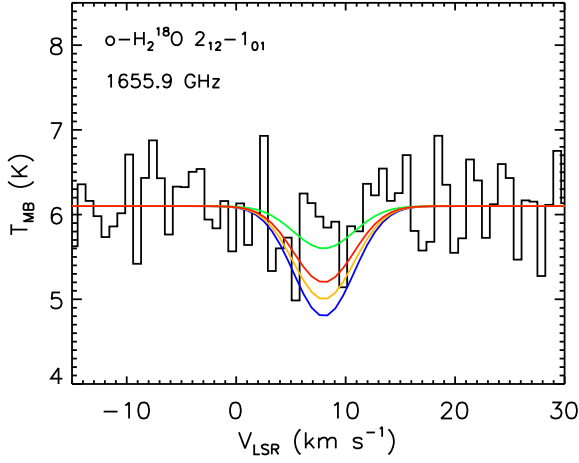}
\caption{Four absorption lines on top of spectra of the ground-state ortho-H$_2$$^{18}$O $2_{12}-1_{01}$ line (black), with continuum in Orion S.
         Assuming that the OPRs are 1, 2, 2.5, and 3, the derived optical depth of 0.08, 0.16, 0.20, and 0.24 are presented, respectively, as green, red, yellow, and blue.
         }
       \label{fig3}
\end{figure}

In Sect.~\ref{orions}, we estimate the intensity of the ground-state line of ortho-H$_2$$^{18}$O $2_{12}-1_{01}$ (1655.9 GHz) assuming that this line appears in absorption to constrain the OPR in Orion S. In Fig.~\ref{fig3} we present four absorption lines on top of the ground-state ortho-H$_2$$^{18}$O $2_{12}-1_{01}$ line. The green, red, yellow, and blue lines represent absorption lines with optical depth of 0.08, 0.16, 0.20, and 0.24, respectively.

\end{appendix}



\end{document}